\documentclass[preprint,prl,showpacs,preprintnumbers,amsmath,amssymb]{revtex4}

\usepackage{graphicx}
\usepackage{dcolumn}
\usepackage{bm}
\usepackage{amsfonts}

\newcommand{\nl}{\nonumber \\ }

\newcommand{\be}{\begin{equation}}  
\newcommand{\ee}{\end{equation}}  
\newcommand{\bear}{\begin{eqnarray}}  
\newcommand{\eear}{\end{eqnarray}}  
\newcommand{\ba}{\begin{array}}  
\newcommand{\ea}{\end{array}}

\newcommand{\tp}{\tilde{\pi}} 
 
\newskip\humongous \humongous=0pt plus 1000pt minus 1000pt

\newif\ifdtup

\def\oldreffmt#1{\rlap{[#1]} \hbox to 2\parindent{}}

\def\figfmt#1{\rlap{Figure {#1}} \hbox to 1in{}}  
  
\def\ie{\hbox{\it i.e.}{}}      
  
\def\eg{\hbox{\it e.g.}{}}      
\def\cf{\hbox{\it cf.}{}}

\def\Tr{\mathop{\rm Tr}}

\def\slash#1{#1\!\!\!/\!\,\,}  
\def\beq{\begin{equation}}  
\def\eeq{\end{equation}}  
\def\bea{\begin{eqnarray}}  
\def\eea{\end{eqnarray}}  
\def\half{\frac{1}{2}}  
  
\def\bq{\begin{quote}}  
\def\eq{\end{quote}}

\def\half{\frac{1}{2}}       

\relax

\newdimen\tdim  
\tdim=\unitlength

\begin{document}

\preprint{FERMILAB-Pub-07/096-T} 
\title{$T$-Parity Violation by Anomalies }

\author{Christopher T. Hill \\ and \\
  Richard J. Hill}

\affiliation{
  {{Fermi National Accelerator Laboratory}}\\
  {{\it P.O. Box 500, Batavia, Illinois 60510, USA}} }

\date{\today}

\begin{abstract}  
 Little Higgs theories often rely on an 
 internal parity (``$T$-parity'') 
 to suppress non-standard electroweak effects or to provide a 
 dark matter candidate.
 We show that such a symmetry is generally broken by anomalies, 
 as described by the Wess-Zumino-Witten term. We study a simple
 $SU(3)\times SU(3)/SU(3)$ Little Higgs scheme where we obtain 
 a minimal form for the topological interactions of a single Higgs field.
 The results apply to more general models, including
 $[SU(3)\times SU(3)/SU(3)]^4$, $SU(5)/SO(5)$,  and
 $SU(6)/Sp(6)$.
\end{abstract}



\maketitle

\section{\bf Introduction}

It is of urgent interest to discover the Higgs boson 
and determine whether it is (i) a fundamental
particle, such as in SUSY theories or the 
Standard Model; (ii) a heavy,
broad resonance, 
such as in TeV-scale dynamical models (like the top-seesaw
\cite{topseesaw});
or (iii) a light composite particle~\cite{Kaplan:1983fs}, 
such as in Little Higgs theories~\cite{nima}
with new dynamics at the $\sim 10$~TeV scale.
Although (i) has
received orders of magnitude more attention
and is correspondingly more refined, 
we note that none of these options is presently ruled out 
by experiment.  In option (iii) the Higgs 
boson  is a (pseudo)
Nambu-Goldstone boson (pNGB) of a spontaneously broken chiral
symmetry. We focus presently on this possibility,
as it is distinguished from (i) and (ii) by remarkable topological
features. In the following we will not distinguish 
between  ``light composite Higgs boson" or ``Little
Higgs boson."

In Little Higgs models some new UV dynamics at $\Lambda\sim 10$ TeV
is imagined to produce a condensate, 
yielding pNGB mesons with decay constants of order 
$F\sim \Lambda/4\pi \sim 1\, {\rm TeV}$. 
Iso-doublet pNGB's, analogues of the $K$-meson in QCD, 
play the role of Higgs scalar fields and develop VEV's at the weak scale, 
$v/\sqrt{2} = 175$ GeV, breaking electroweak (EW) symmetry in the usual way. 
With suitable symmetries forbidding large mass corrections
to the composite Higgs bosons, and a mechanism for generating the EW
symmetry-breaking potential, one hopes to solve the ``little hierarchy''
problem with $\Lambda \sim 10 $ TeV, $F\sim 1$ TeV and
a naturally light Higgs boson at a scale of a few hundred GeV.

If the Higgs is a pNGB there
will generally be ``topological interactions'' that reflect an
anomaly structure of the underlying theory at scale $\Lambda$. 
These interactions, which are suppressed by powers
of $F$, distinguish a chiral meson field from an ordinary field.  
This physics represents the holographic aspect of a chiral lagrangian theory.
Topologically twisted field configurations of a $D=5$ theory 
cast shadows on the $D=4$ surface.
These topological effects are contained in the Wess-Zumino-Witten (WZW) 
term~\cite{Wess:1971yu,Witten,Hill:2007nz}, which must
be included as part of the full effective action.  
The WZW term is a remarkable object and its full implications
are well beyond the scope of the present paper.
For a Little Higgs theory 
from a purely $D=4$ perspective, it contains the full 
anomaly physics of the UV completion
theory, expressed only 
in terms of gauge fields and pNGB's~\cite{BJ,Adler,bardeen}.   
It is specified by an integer quantity, \eg, 
the number of ``colors'' of the constituent ``techni-quarks''.   
The WZW 
interactions of Little Higgs bosons thus 
probe the underlying theory 
above the scale $\Lambda$, much like the $\pi^0\to\gamma\gamma$ interaction
probes short-distance QCD.

Ignoring the WZW term would erroneously miss a significant part of 
the physics of
any chiral lagrangian theory.  For example, 
in many Little Higgs models it is natural to consider an apparent 
new symmetry dubbed ``$T$-parity''~\cite{Cheng:2003ju,Cheng:2004yc}.  
Phenomenological studies suggest that incorporating such a symmetry 
can alleviate various fine tunings.
The lightest $T$-odd particle has been suggested to act as a dark matter
candidate. 

We will show, however, that $T$-parity
is generally violated by anomalies, and hence by
the WZW term. This leads to the decay of the lightest $T$-odd 
particle into gauge fields, and other effects as well.
Indeed, this would also happen
in QCD where the $\pi^0$ can be viewed as a ``$T$-odd'' field,
under which $\pi^0\rightarrow -\pi^0$, while the
photon is ``$T$-even,'' $A_\mu \rightarrow A_\mu$. 
$T$-parity is conserved in the QCD chiral Lagrangian without the 
WZW term, but is violated by anomalies.  For example, the decay of
the ``dark matter candidate'' $\pi^0$ proceeds via $\pi^0\to\gamma\gamma$. 

This is also the 
fate of $T$-parity in Little Higgs theories.
Omitting
the WZW term would lead to the {\em incorrect} 
conclusions that $T$-parity partners must be produced in pairs, and that
the lightest $T$-odd state is stable.  Including it leads
to a rich set of new interactions that can probe the underlying
UV completion of the effective theory at the fundamental
scale $\Lambda$.

The WZW term could in principle be absent, 
but we know of no sensible UV theory in which that would be the case. 
The chiral lagrangian theory will unitarize itself into
something new at scales above $\Lambda$, 
and a QCD-like theory in $D=4$ is the most
straightforward possibility. Alternatively, one might argue that 
the chiral lagrangian becomes a compactified $D\geq 5$ Yang-Mills
theory. 
However, there
is a rich set of topological objects in $D\geq 5$ that
require Chern-Simons terms for their complete description.~%
\footnote{
For example, the instanton becomes a world-line (particle) in $D=5$ and has two
associated conserved topological currents. These currents can only be generated
if Chern-Simons terms are appended to the theory \cite{hillramond}.  
}
Moreover, if the UV theory derives from 
extra dimension(s) we must still provide for the chirality of 
the ordinary quarks and
leptons.  This implies ``chiral delocalization'' in the
extra dimensions, and in turn, Chern-Simons terms that
holographically descend to the WZW term at low energies in $D=4$. 

We will study a simple composite Higgs model that
imitates QCD, with the chiral symmetry breaking pattern 
$SU(3)\times SU(3) \to SU(3)$.    
The Higgs is identified with the kaon.  
The EW $SU(2)\times U(1)$
interactions are a subgroup of the vectorial $SU(3)$ subgroup
and have no internal gauge anomalies. 
We then introduce a $U(1)_5$ axial
vector gauge field, $\tilde{B}$, coupled to the $\lambda^8$ axial
current. 
This situation
imitates what happens in all Little Higgs theories with $T$-parity.
We study the interactions of the Higgs, $W$, $Z$, $\gamma$ 
and the $T$-odd field $\tilde{B}$
in the presence of the WZW term.  

We will see that in a naive treatment there are 
Higgsless ``Chern-Simons'' operators generated by
the WZW term, such as $\tilde{B} Z dZ$, and $\tilde{B} W dW$.~%
\footnote{In a more precise language,  
Chern-Simons operators arise only in odd $D$, as
in $\epsilon^{ABCDE}A_Ad_BA_Cd_DA_E$. Here we use the pejorative meaning
for $D=4$ operators such as 
$\epsilon_{\mu\nu\rho\sigma}A^\mu B^\nu d^\rho B^\sigma$ 
or $\epsilon_{\mu\nu\rho\sigma}\Tr(A^\mu B^\nu B^\rho B^\sigma)$.
}
These operators carry gauge anomalies, and 
are a symptom of uncancelled $\tilde{B}$ gauge 
anomalies in the UV completion theory.
We briefly describe two
methods of cancelling anomalies in the UV
theory, a ``lepton'' sector, and a ``mirror'' sector.
At the level of the WZW term it is easy to implement
the effect of the anomaly cancellation sectors.
When the anomalies are cancelled, the 
Chern-Simons operators drop out, 
as they must by gauge invariance. 
Gauge invariant operators involving the Higgs field remain, and
these represent the universal $T$-parity violating
topological interactions of 
the Higgs, $W$, $Z$, $\gamma$ and $\tilde{B}$.  
As an example, we obtain the
partial decay width of $\tilde{B}\rightarrow ZZ$ 
at order ${\cal{O}}(v^2/F^2)$. This clearly demonstrates that 
$\tilde{B}$ cannot be a dark-matter
candidate in this scheme.

We have thus derived the minimal form of a single Higgs fields participating
in a $T$-parity violating interaction. 
The results can be applied to  multi-Higgs boson extensions, or 
extensions with additional gauge bosons. 
We briefly discuss 
issues relevant to $[SU(3)\times SU(3)/SU(3)]^4$, $SU(5)/SO(5)$,  and
 $SU(6)/Sp(6)$ models.

\section{$\bm{ SU(3)\times SU(3)/SU(3) }$ Little Higgs model} 

Consider a QCD-like theory with strong 
gauge group $SU(N_c)$,
and with $SU(3)$ flavor triplets of techni-quarks, $\Psi_L$ and $\Psi_R$,
that transform in the fundamental representation with $N_c$ colors.  
The strong interaction results in a condensate $\langle \psi_L^i
\bar{\psi}_R^j \rangle \sim \Lambda^3 \delta^{ij}$, leading to an
$SU(3)_L\times SU(3)_R\times U(1) / SU(3)\times U(1)$ chiral
Lagrangian described by the $3\times 3$ unitary matrix field $U^{ij}
\sim \psi_L^i \bar{\psi}_R^j$: 
(we ignore the axial $U(1)$ pNGB, \ie, the $\eta'$)
\be
\label{U}
U =\exp(2i {\tilde{\pi}}/F)\,,  
\quad 
\tilde{\pi}=\sum_{a=1}^8
\pi^a\lambda^a/2 
= 
\frac12 
\left(
\begin{array}{cc} 
\sum_{a=1}^3 \pi^a \tau^a +  \eta \openone_2 / \sqrt{3} & H \\ 
H^\dagger & - 2\eta/  \sqrt{3} 
\end{array}
\right) 
\,.
\ee
The Higgs doublet is identified with the kaon.

At the techni-quark level this system is coupled to
left- and right-handed gauge fields, $A_L$ and $A_R$
respectively, which include the EW
fields and additional new gauge interactions:
\beq
\label{techni}
D_\mu \Psi_L = (\partial_\mu - i A_{L\mu})\Psi_L,
\qquad
D_\mu \Psi_R = (\partial_\mu - i A_{R\mu})\Psi_R.
\eeq
This induces the covariant derivative on $U$:
\beq
D_\mu U = \partial_\mu U -i A_{L\mu} U +iU A_{R\mu} \,.
\eeq
or, in terms of vector fields, $V= \half(A_R + A_L)= V^a T^a$,
and axial vector fields, $V^5= \half(A_R - A_L)= V^{5a}T^a$: 
\be
\label{covar}
D_\mu U = \partial_\mu U -i [V_\mu , U]  + i\{V^5_{\mu},U\},\; \qquad
D_\mu \Psi = (\partial_\mu - i V_\mu - iV^5_\mu\gamma_5)\Psi
\ee
The low-energy theory is specified 
in terms of $U$, $V_\mu$ and $V^5_\mu$.
The axial vector fields eat mesons to acquire mass
while the vector fields remain massless and transform
in the adjoint under the diagonal $SU(3)$ subgroup.

\subsection{$T$-Parity and ``Reconstruction''}

Following the terminology of Cheng and Low~\cite{Cheng:2003ju,Cheng:2004yc} 
we can introduce the concept of
``$T$-parity'', under which the fields transform as
$V\rightarrow +V$, $V^5\rightarrow -V^5$ 
and $\tilde{\pi} \rightarrow -\tilde{\pi}$.  This symmetry
is independent of ``space-parity'', under which 
$(x^0, \vec{x})\rightarrow (x^0,-\vec{x})$,
$\tp \to \tp$, 
$(V_0, \vec{V})\rightarrow (V_0, -\vec{V})$, and 
$(V^5_0, \vec{V}^5) \rightarrow (V^5_0, - \vec{V}^5)$.
Both
$T$-parity and space-parity are symmetries 
of the ``ordinary'' chiral Lagrangian of QCD, 
\be
\label{kinetic}
{\cal L} = -\frac12{\rm Tr}(F_{\mu\nu}^2) + \frac{1}{2}F^2 {\rm Tr}\left( D_\mu U D_\mu U^\dagger 
+ \dots \right) \,,
\ee
where the ellipsis denotes other invariant combinations of   
$U$ and $D_\mu$.  
It is not possible to write a manifestly 
local (\ie, four-dimensional), and 
globally chiral-invariant term that breaks $T$-parity, 
and thus the symmetry would appear to be exact~%
\footnote{
Note that we can also include a matrix $\Omega$ with 
$\Omega^2=1$ into the definition of $T$-parity:
$\tp \to -\Omega \tp \Omega$, {\it etc}.  For example, 
$\Omega= {\rm diag}(1,1,-1)$ will allow the $K$ 
to be defined as scalar, whereas the 
$\pi$ and $\eta$ are pseudoscalar.  As in 
the $SU(3)$ QCD chiral Lagrangian, this 
has no physical consequences until isospin-violating 
interactions are added.    
}.

Any $D=4$ chiral lagrangian theory
can be viewed as a deconstructed gauge theory 
in $D=5$~\cite{decon}.
We can ``reconstruct'' the $D=4$ chiral lagrangian 
of QCD, mapping it into a corresponding $D=5$ Yang-Mills theory
of flavor. 
``$T$-parity'' is then seen to be equivalent to ``KK-mode parity''.
We consider a manifold in $D=5$ with boundary
branes at $y=0$ and $y=R$.  On the manifold
we have a bulk $SU(3)$ (flavor)
Yang-Mills gauge field $B_A = B_A^a\lambda^a/2$. 
If we apply the boundary condition $[D_5, F_{\mu\nu}] =0$
to the gauge fields
on the branes we obtain the spectrum of QCD.
The general configurations of $B_A$ that are allowed on the $y$ interval
are classified in terms of reflections about $y=R/2$. A
$B_\mu$ zero mode, constant in $y$, is  even under this reflection
and is identified with $V_\mu$, and
assigned $T$-parity $+1$ (this corresponds to photons, or other
fundamental vector gauge fields, or to the $\rho$ octet).
The $B_5$ zero-mode, identified with $\tilde{\pi}$, is constant and 
thus also even under
reflection about $y=R/2$, but is the 5th
component of a 5-vector and thus assigned $T$-parity $-1$.  
The first KK-mode of $B_\mu$ is identified with 
$V^5_{\mu}$.  It has wave-function 
 $\cos(\pi y/R)$ (the mode $\sin(\pi y/R)$ is forbidden
by the b.c.'s) and is
odd under reflection about $y=R/2$, and thus assigned $T$-parity
$-1$. This mode eats the first $\sin(\pi y/R)$ KK mode of $B_5$
which is also $T$-parity even. The first physical KK-mode of $B_5$
(corresponding to the $a^0$ octet) has wave-function
$\cos(\pi y/R)$  and is thus assigned $T$-parity of $+1$, and so forth.  

$T$-parity is, however, not a good symmetry of QCD. 
For example,  
$\pi^0 \to \gamma \gamma$ is an allowed process, and 
we also have that the $\phi$ meson (the $V^8$ in our notation) decays both 
to $K\bar{K}$ and to $\pi\pi\pi$. There is no 
assignment of a conserved $T$-parity
that is consistent with these facts.  
The resolution is that we must incorporate the 
WZW term into the effective action.
The WZW term is four-dimensional and 
globally chiral invariant, although
not manifestly so.  
For general gauge fields, it is not  
gauge invariant, reflecting the anomaly structure of the underlying QCD
theory.  However, when only non-anomalous (\eg, vector) 
generators are gauged, it is gauge invariant, albeit again not 
manifestly so.    
The WZW term is odd under independent space-parity or $T$-parity reflections, 
and only the combination of these two parities 
survives as the true parity symmetry.  

From the perspective of the $D=5$ reconstructed theory,
the WZW term is seen to arise from the Chern-Simons
term which involves
$\epsilon^{ABCDE} B_A\partial_B B_C \partial_C B_D + ...$.
Under compactification the Chern-Simons term resolves into two terms, 
$\sim B_5\epsilon^{\mu\nu\rho\sigma}\partial_\mu B_\nu
\partial_\rho B_\sigma$  and 
$\sim \epsilon^{\mu\nu\rho\sigma}B_\mu \partial_5 B_\nu
\partial_\rho B_\sigma$.  Both terms break ``KK-mode parity''
while conserving overall $D=5$ parity. 
Again, the low-energy effective action contains only
a single overall parity symmetry. 
We remark that {\em any higher dimension theory}, \eg, 
Randall-Sundrum models,
with $B_5$ as a Higgs field and with chiral fermion
delocalization, will have CS term effects.
Moreover, a theory with $\Psi_L$ ($\Psi_R$)
fermions on the $y=0$ ($y=R$) brane, requires the 
Chern-Simons term for anomaly cancellation. The full WZW
term arises from the Chern-Simons
term and the Dirac determinant when we integrate out the 
fermions.

Let us focus on the gauge/Higgs sector of 
the model, ignoring the ordinary standard model quarks and leptons, 
and the mechanism that gives rise to the Higgs potential.    
Consider gauging the $SU(2)\times U(1)$ subgroup of 
the vectorial $SU(3)$, and a $U(1)_5$ axial
subgroup of $SU(3)\times SU(3)$.
We specify this by the covariant derivative of Eq.(\ref{covar}) and
the gauge fields:
\begin{equation}
\label{Afield}
V_\mu = \sum_{a=1}^3 g_2 W_\mu^a \lambda^a/2 
-  g_1 B_\mu \lambda^8/2 \sqrt{3}
\qquad
V^5_{\mu} = \tilde{g} \tilde{B}_\mu\lambda^8/2 \,.
\end{equation} 
The axial vector field $\tilde{B}$
eats the $\eta$ and becomes massive, with a mass:
\beq 
m_{\tilde{B}}=\tilde{g}F
\eeq
The Higgs field transforms as an isodoublet, with covariant derivative
\beq
\label{Higgscov}
D_\mu H = \partial_\mu H - i g_2 W_\mu^a\frac{\tau^a}{2}H + \frac{i}{2} g_1
B_\mu H
\eeq
The $T$-parity conserving couplings of $\tilde{B}$ in unitary gauge (in which
it eats the $\eta$) can
be inferred from the full Eq.(\ref{kinetic}).

This gauging is anomaly-free with respect to the color gauge group, 
but introduces $SU(2)^2\times U(1)_5$ and various $U(1)_5 U(1)^{2}$ 
and $U(1)_5^{3}$ anomalies. 
These anomalies can be cancelled either by a spectator ``lepton'' sector, 
or by introducing a ``mirror sector'' with the opposite chirality, 
as we discuss in the following section.

The WZW action can be evaluated straightforwardly as in QCD.
We use the form of Kaymakcalan, Rajeev and Schechter~\cite{kay}
(their Eq.(4.18)). 
Relevant issues for adapting this
to Little Higgs theories are described in our earlier 
paper~\cite{Hill:2007nz}.
We work in unitary gauge for the heavy fields,
where $\tilde{B}$ eats $\eta$.  
Through order $1/F^2$, 
we find:
\bea
\label{WZsu3}
\Gamma_{WZW} & = & \int d^4x\; \frac{\tilde{g} N_c}{24\sqrt{3}\pi^2}
\epsilon^{\mu\nu\rho\sigma}  \tilde{B}_\mu \big[
\nonumber \\ & & \qquad
 -\frac{1}{3}g_1^2[B_\nu \partial_\rho B_\sigma]  
 + 2{g_2^2}\Tr[W_\nu \partial_\rho W_\sigma] - \frac{3ig_2^3}{2}\Tr[W_\nu W_\rho
 W_\sigma] 
 \nonumber \\
& & \!\!\!\!\! \!\!\!\!\! \!\!\!\!\!
-\frac{ig_1}{4F^2} F^B_{\nu\rho}[H^\dagger ({D}_\sigma H) - (D_\sigma H^\dagger) H]
- \frac{ig_2}{F^2} [H^\dagger F^W_{\nu\rho} (D_\sigma H)  - (D_\nu H^\dagger) 
F^W_{\rho\sigma} H] \big] \,, 
\eea
where $DH$ is given in Eq.(\ref{Higgscov}). 
Here $F_{\mu\nu}^W$ and $F_{\mu\nu}^B$ are field strengths for the $B$,
$F^B_{\mu\nu} = 2\partial_{[\mu}B_{\nu]}$, 
and $W$: [Square brackets around indices denote antisymmetrization, 
$A_{[\mu}B_{\nu]} \equiv \half (A_\mu B_\nu - B_\mu A_\nu )$ ]
\beq
F^W_{\mu\nu} =
 \left(\begin{array}{cc} \partial_{[\mu} W^3_{\nu ]} -
 ig_2W^+_{[\mu}W^-_{\nu ]} & 
\sqrt{2}(\partial_{[\mu} W^+_{\nu ]} - ig_2W^3_{[\mu}W^+_{\nu ]}) \\ 
\sqrt{2}(\partial_{[\mu} W^-_{\nu ]} + ig_2W^3_{[\mu}W^-_{\nu ]}) & 
-\partial_{[\mu} W^3_{\nu ]} + ig_2W^+_{[\mu}W^-_{\nu ]}
\end{array}\right )
\eeq
where
\beq 
{W}_\mu^3 =  Z_\mu^0 \cos\theta_W + A_\mu \sin\theta_W \qquad
{B}_\mu = - Z_\mu^0 \sin\theta_W  + A_\mu \cos\theta_W  \,. 
\eeq 
Here $A_\mu$ ($Z^0_\mu$) is the physical photon ($Z^0$) vector potential.

However, any
action containing  Eq.(\ref{WZsu3}) alone, cannot be physically correct.  
The terms 
of Eq.(\ref{WZsu3}) containing $\tilde{B}W \partial W$ and $\tilde{B}B \partial B$ 
generate anomalies and describe the disallowed 
decay of a massive spin-1 field into
two massless spin-1 fields
in violation of gauge invariance and 
the Landau-Yang theorem  (the $\rho$ cannot
decay to two photons!). 
This is a symptom of the fact that the 
axial $\lambda^8$ symmetry of $\tilde{B}$ 
is anomalous at the level of our
fundamental techni-quark theory in Eq.(\ref{techni}).
A consistent theory requires additional 
anomaly-cancelling structure, and this structure will modify
the decay amplitudes. We will see that it is 
easy to represent this in a fairly general way.

\subsection{Anomaly cancellation}

A ``lepton sector'' can be constructed that cancels
gauge anomalies and makes the model consistent.  Consider, \eg,
$N_c$ ``leptons'' with covariant derivative:
\begin{equation}
\label{leptons}
D_\mu L = 
\partial_\mu L - iV_{\mu} L + iV{}^5_\mu\gamma^5 L \,.
\end{equation} 
Here we have flipped the sign of the $V{}^5_\mu$ interaction
for the leptons relative to the techni-quarks 
in Eq.(\ref{covar}).
We postulate that the leptons acquire mass via their own Higgs mechanism, 
with the gauge invariant Lagrangian: 
\be 
\Delta {\cal L} = \half f^2 {\rm Tr}( D_\mu U^\prime D_\mu U^{\prime \dagger} ) 
+ \bar{L} i\slash{D} L 
- m_L \left( \bar{L}_L U^\prime L_R + h.c. \right)  \,, 
\ee
and here we must introduce an ``axion'' field, $a$ to
preserve the axial $\lambda^8$ symmetry: 
\be 
U^\prime = e^{ 2 ia \lambda^8/f } \,. 
\ee
Under $U(1)_5$ we have $L_R\rightarrow e^{i\theta\lambda^8}L_R$, 
$L_L\rightarrow e^{-i\theta\lambda^8}L_L$ and
the axion and $\tilde{B}$ transform as: 
\be
\tilde{g} \tilde{B}_\mu \rightarrow \tilde{g} \tilde{B}_\mu  
+ \partial_\mu\theta ,
\qquad \qquad
a/f \to a/f - \theta. 
\ee
With the leptons and techni-quarks the theory is now anomaly free.
Integrating out the leptons in a large mass
limit will give a new WZW action, 
\be
\Gamma_{WZW} = \Gamma_{WZW}(U,\tilde{B},W,B) 
- \Gamma_{WZW}(U^\prime,\tilde{B},W,B) \,. 
\ee
The relative minus sign comes from flipping the $\tilde{B}$
coupling constant sign in Eq.(\ref{leptons}).

Note that integrating out the leptons
is a convenience 
in that they would otherwise complicate the computation
of physical processes. 
If the leptons were light we would still
have overall anomaly cancellation, but we would have to compute
open lepton processes and add them to WZW processes of the techni-quarks.  

With anomaly cancellation in place
we see immediately that
the offending Chern-Simons terms, 
 $\tilde{B} W \partial W$ and
 $\tilde{B} B \partial B$, 
cancel in the sum of WZW terms.
There remain terms that involve the Higgs fields
of the form  $\sim H^\dagger H \tilde{B} W \partial W$ 
 and $\sim H^\dagger H \tilde{B} B \partial B$. 
The scalar axion field $a$ remains in the low-energy spectrum, and
is neutral under EW $SU(2)\times U(1)$.  
Since we are not interested in the fate of the 
axion and the $\eta$, or the $\pi$ pNGB's, 
presently we can simply set them to zero 
and 
generate the physical amplitudes of interest
using the formula:
\be
\label{simp}
\Gamma_{WZW} \approx \Gamma_{WZW}(U,\tilde{B},W,B) 
- \Gamma_{WZW}(1,\tilde{B},W,B) \,.
\ee

An alternative scheme that doesn't involve leptons or a fundamental
axionic pseudoscalar field 
can be constructed by supposing the mirror fermions are themselves
coupled to a strong gauge force, causing a mirror fermion 
condensate. 
Gauging both sectors in an identical manner, 
\begin{align} \label{U12}
U_1 & \sim \Psi_{1L} \bar{\Psi}_{1R} 
\to e^{i\epsilon_L} U_1 e^{-i\epsilon_R} \,, \nl
U_2 & \sim \Psi_{2R} \bar{\Psi}_{2L} 
\to e^{i\epsilon_L} U_2 e^{-i\epsilon_R} \,, 
\end{align} 
again ensures that all Chern-Simons terms and gauge
anomalies cancel between 
the two sectors.  The low-energy theory becomes 
a two-Higgs doublet model, 
with the anomaly physics described by 
$\Gamma_{WZW}(U_1)- \Gamma_{WZW}(U_2)$. 

\section{The Physical T-Parity Violating WZW Interaction}

Including the minimal anomaly cancelling spectators
discussed above, and neglecting additional pNGB's, 
we see that the Chern-Simons terms cancel and
we are left with the physical
WZW term interaction of a single Little Higgs doublet
with $\tilde{B}$ through order $1/F^2$:
\bea
\label{WZsu32}
\Gamma_{WZW} & = & \int d^4x\; \frac{\tilde{g} N_c}{24\sqrt{3}\pi^2}
\epsilon^{\mu\nu\rho\sigma}  \tilde{B}_\mu \big[
\nonumber \\
& & \!\!\!\!\! \!\!\!\!\! \!\!\!\!\!
-\frac{ig_1}{4F^2}F^B_{\nu\rho}[H^\dagger ({D}_\sigma H) - (D_\sigma H^\dagger) H]
- \frac{ig_2}{F^2} [H^\dagger F^W_{\nu\rho} (D_\sigma H)  - (D_\nu H^\dagger) 
F^W_{\rho\sigma} H] \big] \,. 
\eea 
Note the pair of operators that are generated at this level
of the form $\epsilon^{\mu\nu\rho\sigma}\tilde{B}_\mu
[H^\dagger F_{\nu\rho}({D}_\sigma H) - (D_\nu H^\dagger) F_{\rho\sigma}H]$.
This is a generic operator structure for gauge/Higgs interactions
in the WZW term
for all Little Higgs theories containing a tree-level
$T$-parity. 

Since the theory is now anomaly free we can pass to unitary gauge
whence the Higgs field takes the form:
\beq
H = 
\frac{1}{\sqrt{2}} \left(\begin{array}{c} v + h^0  \\ 
0
\end{array}\right ) \,, 
\eeq
with $v = 246$ GeV.
The interaction takes the explicit form:
\bea
\label{final}
 \Gamma_{WZW} & = & \frac{-\tilde{g}g^2_2N_c }{96\sqrt{3}\pi^2 F^2} 
\int d^4x\,( v + h^0 )^2 \epsilon^{\mu\nu\rho\sigma} \tilde{B}_{\mu} \times
\nonumber \\
&&
\left[ 
2 \sqrt{1+ \tan^2\theta}
\left(\partial_{\nu}  Z_{\rho}^0 \cos\theta 
+ \partial_{\nu}A_{\rho} \sin\theta
-ig_2W^+_{\nu}W^-_{\rho} \right)
 Z_\sigma^0 \right.
 \nonumber \\
& &  
+\left.
2\left[ 
(D^A_{\nu} W^+_{\rho}) 
{W}_\sigma^- 
+ 
(D^A_{\nu} W^-_{\rho})
{W}_\sigma^+
\right] - 4 i g_2 \cos\theta Z^0_\nu W^+_\rho W^-_\sigma\right.
 \nonumber \\
& &  
 \left. - \tan\theta\sqrt{1 +\tan^2\theta}
\left( \partial_\nu Z^0_\rho \sin\theta  
- \partial_\nu A_\rho \cos\theta\right)Z_\sigma^0 
\right] 
\eea 
where $D^A_{\nu} W^\pm_{\nu } = (\partial_\mu \mp ieA_\mu)W^\pm_\nu$
and $A_\mu$ is the photon vector potential.
Note that Eq.(\ref{final}), which is written in unitary
gauge for the $\tilde{B}$, $W$, and $Z$, is manifestly invariant under
electromagnetic gauge transformations.

\subsection{Brief Survey of Physical Processes}

Physical processes described by Eq.(\ref{final}) all violate
$T$-parity (and space-parity), 
conserving overall parity.  Some examples worthy
of study for future colliders include:
\bea
e^+e^- \; \makebox{or}\; q\bar{q} \; \makebox{or}\; \mu^+ \mu^-
& \rightarrow & (\gamma^*, Z^*) \rightarrow \tilde{B}+Z; 
\; \tilde{B} + \gamma; \; \tilde{B} + WW
\nonumber \\
e^+e^- \; \makebox{or}\; q\bar{q} \; \makebox{or}\; \mu^+ \mu^- 
& \rightarrow & (\gamma^*, Z^*) \rightarrow \tilde{B}+h^0; 
\; \tilde{B} + 2h^0
\nonumber \\
q\bar{q} & \rightarrow & W^* \rightarrow \tilde{B}+W; \; \tilde{B} + W+ h^0;
\; \tilde{B} + W+ 2h^0
\eea
These processes could in principle be used 
to measure $N_c$ for the UV completion theory.
They are, however, suppressed  in rate by $N_c^2/F^{4}$ and would probably
be best suited for a very luminous ILC, but are unlikely to
be observable for large $F$.  We have not computed
the cross-sections, and suspect they are quite small
since they are effectively loop-level.

An interesting possibility is that $\tilde{B}$ couples to
the electron, muon, or quarks directly and can be produced in
the $s$-channel. Then we have interesting
processes such as:
\bea
e^+e^- \; \makebox{or}\; q\bar{q}\; \makebox{or}\; \mu^+ \mu^-
& \rightarrow & (\gamma^*, Z^*,\tilde{B}^*) \rightarrow 
\nonumber \\
& & \!\!\!\!\!  \!\!\!\!\! \!\!\!\!\! \!\!\!\!\!  \!\!\!\!\! \!\!\!\!\!
\tilde{Z}+Z + (0,1,2)h^0; 
\; \tilde{Z} + \gamma + (0,1,2)h^0; \;  WW + (0,1,2)h^0 \; \makebox{,etc.}
\eea 
Processes such as $e^+e^-\rightarrow\tilde{B}^* 
\rightarrow ZZ$ could interfere against normal EW
physics, such as $e^+e^-\rightarrow ZZ$ producing 
an interference term that scales as $N_c/F^2$
with a chance at observability in detailed angular correlation studies
at an ILC, CLIC or muon collider.  
Polarization may be a useful attribute to study in such processes.

Finally, as we have emphasized, even if $\tilde{B}$ has
no direct coupling to light fermions, it will necessarily
decay through the $T$-parity violating processes:
\bea
 \tilde{B} & \rightarrow & 
\tilde{Z}+Z + (0,1,2)h^0; 
\; \tilde{Z} + \gamma + (0,1,2)h^0; \;  WW + (0,1,2)h^0 \; \makebox{,\, etc.}
\eea
As an explicit application, we compute the partial width:
\be\label{Bhatwidth}
\Gamma(\tilde{B}\to ZZ) \approx {1\over 2\pi} 
\left( \tilde{g}^3 N_c \over 144 \pi^2 \right)^2 
{m_Z^2\over m_{\tilde{B}}} \,, 
\ee
to leading order in $\sin \theta_W$.  Here we have 
used the relation $m_{\tilde{B}}= \tilde{g}F$ to 
simplify the result. 

We have ignored 
the $\pi$ and axion-$\eta$  
relic pNGB's which remain in the physical spectrum. These will also have
associated
anomalous interactions, such as $\pi^0\to VV$, $a' \to VV$, where 
$V$ is a vector boson.
In extensions of this minimal model, we could gauge, \eg, two copies 
of $SU(2)$, such that $\pi$ would be eaten by a heavy $\tilde{W}$ gauge
boson.   Note that the pNGB's correspond to axial generators and 
are always odd under $T$ parity, 
while $B$ and $W$ correspond to vector generators and are therefore even. 
That $\tilde{B}$ and $\tilde{W}$ have a definite transformation (odd) 
under $T$ parity relies on a further assumption of equality between 
coupling constants, $g_L=g_R$.   

\section{Popular Little Higgs Models} 

We presently survey a number of models in the literature for 
which an apparent $T$-parity can be defined.   Most of these models
are incomplete, and 
additional anomaly-cancelling structure is necessary for consistency.  

The gauge structure of such a model must be 
sufficiently rich to leave an 
unbroken $SU(2)\times U(1)$ symmetry to be identified with 
the EW interactions.
A model that incorporates one-loop cancellation 
of radiative corrections to the Higgs mass~\cite{nima} also    
necessarily gauges broken $U(1)$ (and $SU(2)$) generators.  
The same phenomenon occurs in more general 
composite models designed to break EW symmetry 
by vacuum misalignment~\cite{Kaplan:1983fs}.  
Since the strong dynamics prefers 
a vacuum orientation that preserves EW symmetry, the misalignment 
can be achieved only if a broken $U(1)$ generator is gauged with 
sufficient strength.  
These broken $U(1)$ symmetries are, in general, anomalous, and 
entail the existence of an additional sector for anomaly cancellation. 

The symmetry-breaking pattern $SU(n_f)\times SU(n_f)\to SU(n_f)$
is expected for a condensate of $2n_f$ Weyl fermions, 
with $n_f$ of these in the fundamental representation of a strong 
color group $SU(N_c)$, and $n_f$ in the anti-fundamental.  $n_f=3$ is
the smallest value for which the ``Little-Higgs cancellation'' of one-loop
mass corrections can be 
implemented.  The existence of the WZW term can be traced to the 
nontrivial homotopy group for $SU(n)\times SU(n)/SU(n) = SU(n)$: 
$\pi_5(SU(n)) = \bm{Z}$ (for $n\ge 3$).      
The conclusions drawn from the $SU(3)\times SU(3)\to SU(3)$ model 
apply more generally.  We mention here some specific examples.  

\subsection{ $\bm{ SU(5)/SO(5)} $  }  
The symmetry breaking pattern $SU(n_f)\to SO(n_f)$ 
is expected for a condensate of $n_f$ Weyl 
fermions in a real representation of
a strong color group.  $n_f=5$ is the smallest value for which the
Little-Higgs cancellation can be implemented.   We note that 
$\pi_5(SU(N)/O(N))= \bm{Z}$ (for $n\ge 3$).  
This theory is described by a chiral field $\Sigma$ transforming as
\be 
\Sigma \to e^{i\epsilon} \Sigma e^{-i R(\epsilon)} \,, 
\ee 
where $R(t^a) = \pm t^a$ for the unbroken and broken generators, 
respectively.~%
\footnote{
Alternatively, the low-energy theory can be arrived at by
embedding inside of $SU(5)\times SU(5)/SU(5)$ with the extraneous NGB's 
removed by a strongly-coupled gauge field. 
} 
The WZW term for this model was derived in \cite{Hill:2007nz} and is 
similar to (\ref{WZsu3}). 
Again, the action is unacceptable in isolation and
requires an anomaly cancelling sector.  
This will again cancel the pure 
$\tilde{B}W \partial W$ and $\tilde{B} B \partial B$ terms,
leaving allowed  
$\tilde{B}H^\dagger H W \partial W$ and $\tilde{B} H^\dagger H B \partial B$
terms~\cite{Hill:2007nz}.  
The  same general arguments apply here as in the previous case: 
$T$-parity is violated, the model is incomplete
without an anomaly cancelling sector, and $\tilde{B}$ is unstable.

Note that $\pi_3(SU(5)/SO(5)) = Z_2$, and the theory must contain
a skyrmion.  The skyrmion reflects ``baryons'' at the
scale $\Lambda$, and dictates the need for the WZW term to generate
the corresponding Goldstone-Wilczek current.  

\subsection{ $\bm{ SU(6)/Sp(6) } $ } 

The symmetry breaking pattern $SU(n_f)\to Sp(n_f)$, for $n_f$ even, 
is expected for a condensate of fermions in a pseudo-real representation of
a strong color group.  $n_f=6$ is the smallest value for which the
Little-Higgs cancellation can be implemented~\cite{Low:2002ws}.  
Again we note that
$\pi_5(SU(n)/Sp(n))=\bm{Z}$ (for $n\ge 4$).  
The low energy theory is a two-Higgs doublet model.  We do not pursue further
details here, but note again that a
$T$-parity can be defined to act such that
unbroken and broken generators are even and odd under
$T$-parity.  The WZW term is odd under $T$-parity, 
and mediates transitions
between $T$-even and $T$-odd states.~%
\footnote{
Note that in contrast to the $U(1)$ charges advocated in Ref.~\cite{Low:2002ws}, 
it is possible to choose generators that do not induce additional 
$U(1)$/color anomalies --- \eg, in the coordinates of Ref.~\cite{Low:2002ws}, 
$Y_1 \propto (1,1,1,1,1,-5)$, and 
$Y_2 \propto (1,1,-5,1,1,1)$. 
}

It is interesting that, while the WZW term exists here, there is
evidently no skyrmion, since $\pi_3(SU(6)/Sp(6)) = 0$.  A similar situation
arises in the Kaplan-Schmaltz models~\cite{Hill:2007nz} where
we cannot construct a Goldstone-Wilczek current.

\subsection{ $\bm{ [SU(3)/SU(2)]^2 } $ } 
It is possible to search for an alternate definition of $T$-parity.   For
example in Eq.(\ref{U12}), we could consider 
$U_1 \leftrightarrow U_2$.  This is a symmetry of the chiral Lagrangian
when the WZW term is omitted, but is broken once it is included.  This fact
simply reflects
the underlying chirality of the fermions: 
$A_L$ is coupled to a left-handed 
fermion in $U_1$, but to a right-handed fermion in $U_2$.  
The same conclusions hold in various limiting cases.  For example, strongly 
gauging a full $SU(3)_R$ in (\ref{U12}) results in the $[SU(3)/SU(2)]^2$ 
Kaplan-Schmaltz theory~\cite{Kaplan:2003uc}; 
explicit $T$-parity violating interactions for this case were 
derived in Ref.~\cite{Hill:2007nz}.

\subsection{ $\bm{ [SU(3)\times SU(3)/SU(3)]^4 } $ }
 
Consider the extension of (\ref{U12}) to a situation with {\it four} 
distinct condensates, 
\begin{align}\label{4U}
U_1 & \sim \Psi_{1L} \bar{\Psi}_{1R} 
\to e^{i\epsilon_L} U_1 e^{-i\epsilon_R} \,, \nl
U_2 & \sim \Psi_{2R} \bar{\Psi}_{2L} 
\to e^{i\epsilon_L} U_2 e^{-i\epsilon_R} \,, \nl
U_3 & \sim \Psi_{3L} \bar{\Psi}_{3R} 
\to e^{i\epsilon_L} U_3 e^{-i\epsilon_R} \,, \nl
U_4 & \sim \Psi_{4R} \bar{\Psi}_{4L} 
\to e^{i\epsilon_L} U_4 e^{-i\epsilon_R} \,.
\end{align}
For example, the model of \cite{moose} 
is of this form, with $\epsilon_L$ generating a full $SU(3)$, 
and $\epsilon_R$ generating $SU(2)\times U(1)$.   
The implementation of $T$-parity in a variant of this model via 
$L\leftrightarrow R$ 
as proposed in \cite{Cheng:2004yc} 
suffers the same fate as the models already 
considered.  
However, the interchange symmetries $U_2 \leftrightarrow U_4$ 
(or $U_1 \leftrightarrow U_3$) are potentially valid symmetries of
the full action
\footnote{
The mechanism of \cite{moose} used to stabilize a flat direction in the 
Higgs potential breaks the $U_2 \leftrightarrow U_4$ symmetry. 
}.
We remark that this model has skyrmion solutions as well.

\section{Discussion}

Composite Higgs models may provide a plausible explanation of 
EW symmetry breaking. 
Anomaly considerations have a crucial impact on the physics.
Far from being a nuisance, the anomaly interactions of composite or 
Little Higgs theories provide a pathway to underlying UV 
physics that is accessible
at the ``low'' energies that will be probed in the next generation 
of colliders.  
Perhaps the most dramatic effects are seen in the decays of new 
heavy particles that would otherwise be stable.
This phenomenon is exactly analogous to the decay of $\pi^0\to\gamma \gamma$ 
which specifies $N_c=3$ in the QCD chiral Lagrangian.  

When the NGB of interest is the kaon, considerations of conserved 
$SU(2)_W$ isospin symmetry (relatedly, strangeness) imply that the anomaly 
interactions are more difficult to observe.  Single $H$ interactions are forbidden
by isospin, while two-$H$ interactions with only massless gauge bosons are forbidden
by parity.  Thus nontrivial interactions start with two Higgs fields, plus either 
additional NGB's, or massive gauge fields.   Nonetheless, such interactions can 
have important implications, \eg, giving rise to the 
$T$-parity violating decay mode studied in (\ref{Bhatwidth}). 

The mechanism of collective symmetry breaking encoded in many such models 
leads to a doubling of certain degrees of freedom, and it is natural to
consider whether a new ``T'' parity can be defined as an exact symmetry.  
We described the general mechanism by which $T$-parity is violated in a
Little Higgs model.   
It may be noted that although the discussion has been framed in terms of 
an underlying fermion UV completion, similar considerations hold for any 
UV completion in which the WZW term does not vanish.   

In what situation can an exact $T$-parity be defined?   
One could consider 
flavor symmetry-breaking patterns, such $SO(n)\times SO(n)/SO(n)$, 
for which the flavor symmetry representations are real, 
and no anomalies occur.  
However, we do not know of any reasonable UV theories 
where such a symmetry-breaking pattern arises---certainly 
not in a theory of strongly-interacting fermions.  
Even in familiar cases such as $SU(n)\times SU(n)/SU(n)$, it is a
logical possibility that the integer coefficient of the WZW term is zero, 
but we do not know what UV completion could possibly lead to this.
Reconstruction suggests that $N_c=0$ would be a property of the 
equivalent $D=5$ Yang-Mills theory, \ie, the Chern-Simons
term is absent for that theory.  Even in the absence
of fermions, such a theory contains instantonic
solitons \cite{hillramond} (\ie, the Euclidean $D=4$
instanton is a world-line) and these match onto 
the Skyrmion under compactification.
Their currents and various static properties, 
like spin and statistics, are controlled
by the Chern-Simons term, which matches onto the WZW term. $D=5$
Yang-Mills does require a UV completion, but to have $N=0$ would
require a suppression of all topological aspects of that theory
by the completion theory.
 
Alternatively, 
$T$-parity can be defined as an exchange symmetry between sectors 
with identical chiral fermion content, \cf\, the discussion after 
Eq.(\ref{4U}).  Whether this symmetry can be maintained after 
adding standard model fermions, and terms generating a Higgs potential,
remains to be explored.  

It is important to 
work out the phenomenological implications of broken $T$-parity; 
in particular, 
missing energy signals at colliders arising from 
a stable $\tilde{B}$ should be 
reconsidered~\cite{Carena:2006jx}.  
Certainly predictions of a dark matter candidate based on 
naive $T$-parity need to be 
revised~\cite{Birkedal-Hansen:2003mp,Birkedal:2006fz}. 
Independently of whether an approximate or exact $T$-parity can be found, 
it is interesting to look for observable effects of the spectator 
lepton sector that is necessary to 
cancel the anomalies of a general 
Little Higgs model.  
Understanding the spectator sector is an important problem.  
The interplay of this sector with the strong interactions 
could in itself
provide a new mechanism for causing vacuum misalignment, and EW 
symmetry breaking. 

We emphasize that these considerations will generally
apply to any models of extra dimensions 
where Chern-Simons terms appear ({\it cf} the second paper
in the sequence of Ref.~\cite{hillch}).  For example,
chiral delocalization appears in Randall-Sundrum
schemes such as the ``composite top" models~\cite{Agashe}.
The CS term must be included and will involve gravitational
as well as gauge interactions.  We should also revisit the question
of when the lightest KK-mode in higher dimensional models
is stable in the presence of a CS term~\cite{Servant}.

\vskip .1in
\noindent
{\bf Acknowledgements}
\vskip .1in
This work was hosted in the Fermilab Theoretical Physics
Department
operated by Fermi Research Alliance, LLC  under Contract No. 
DE-AC02-07CH11359 with the United States Department of Energy.


\end{document}